\documentclass[final,5p,times,twocolumn]{elsarticle}

\usepackage{amssymb}
\journal{Physics Letters B}

\begin{document}
\begin{frontmatter}



\title{Charged Balanced Black Rings in Five Dimensions}


\author{Burkhard Kleihaus, Jutta Kunz and Kirsten Schn\"ulle}

\address{Institut f\"ur Physik, Universit\"at Oldenburg, D-26111 Oldenburg, Germany}

\begin{abstract}
We present balanced black ring solutions of pure Einstein-Maxwell theory in five dimensions. The solutions are asymptotically flat, and their tension and gravitational self-attraction are balanced by the repulsion due to rotation and electrical charge. Hence the solutions are free of conical singularities and possess a regular horizon which exhibits the topology $S^1\times S^2$ of a torus. We discuss the global charges and the horizon properties of the solutions and show that they satisfy a Smarr relation. We construct these black rings numerically, restricting to the case of black rings with a rotation in the direction of the $S^1$.
\end{abstract}


\end{frontmatter}

% \linenumbers

\section{INTRODUCTION}
\label{sec:Introduction}
In four-dimensional spacetime, asymptotically flat electrovac black hole solutions are represented by the static Schwarzschild and Reissner-Nordstr\"om black holes as well as the rotating Kerr and Kerr-Newman black holes. These solutions possess a spherical horizon topology and are uniquely characterized by their global charges, i.e.~their mass, angular momentum and charge \citep{Tangherlini:1963bw}.\\
In the context of string theory and brane world models, higher-dimensional black holes have received much interest in the recent years, opening up the possibility of direct observation in future high energy collisions. The presence of extra dimensions can affect the properties of black holes dramatically. Concerning black holes with a spherical horizon topology, \citet{Tangherlini:1963bw} has derived the counterpart to the Schwarzschild solution, whereas \citet{Myers:1986un} have obtained rotating vacuum black holes. Moreover, a static solution of pure Einstein-Maxwell theory has been found by \citet{Tangherlini:1963bw} and \citet{Myers:1986un}. Higher-dimensional electrically charged stationary solutions are so far only known in closed form for some low energy effective actions related to string theory. However electrically charged rotating black holes have been derived perturbatively by \citet{Aliev:2004ec} in lowest order in the electric charge, and by \citet{NavarroLerida:2007ez} and \citet{Allahverdizadeh:2010xx} up to fourth order. In addition, numerical solutions for the pure Einstein-Maxwell case have been constructed \citep{Kunz:2005nm}.\\ 
A fascinating development was the discovery of the black ring as a topologically different solution of the five-dimensional Einstein equations by \citet{Emparan:2001wn, Emparan:2001wk}. This is a black hole with the horizon topology $S^1\times S^2$ of a torus -- hence the name black ring. For the static case, the solution always suffers from the presence of a conical singularity, whereas balanced solutions with a regular horizon can be obtained for the stationary case, the tension and gravitational self-attraction of the ring being balanced by its centrifugal repulsion. Within a certain parameter range, spherical black holes and black rings coexist, so black hole uniqueness is violated in five dimensions.\\ 
Numerous explicit examples of black rings have been found until now (see e.g.~\citet{Emparan:2006mm, Emparan:2008eg} and references therein). 
Concerning electrically charged black rings, static solutions have been discovered by \citet{Ida:2003wv} in Einstein-Maxwell theory, while \citet{Kunduri:2004da} and \citet{Yazadjiev:2005hr} have derived static solutions in Einstein-Maxwell dilaton theory. \citet{Elvang:2003yy} has obtained a rotating, electrically charged solution in the low-energy limit of heterotic string theory via a Hassan-Sen transformation. Similar to the case of black holes with a spherical horizon topology, a stationary black ring solution of pure Einstein-Maxwell theory is not known in closed form yet. However \citet{Ortaggio:2006ng} have obtained charged rotating black rings perturbatively for small charges.\\ 
% A supersymmetric black ring in five-dimensional minimal supergravity has been obtained by \citet{Elvang:2004rt}.  Further, a rotating, magnetically charged dipole black ring has been found by \citet{Emparan:2004wy}.\\
We here consider numerically obtained stationary, asymptotically flat black ring solutions of pure Einstein-Maxwell theory in five dimensions, which possess a regular event horizon of topology $S^1\times S^2$ and exhibit a rotation in the direction of the $S^1$. We investigate the physical properties of these black objects. The mass $M$, the angular momentum $J$, the electric charge $Q$ as well as the magnetic moment $\mathcal{M}_{\varphi}$ are determined from an asymptotic expansion. We show that the gyromagnetic ratio deviates from the perturbatively found value $g=3$ for higher values of the charge. The Hawking temperature $T_H$ and the horizon area $\mathcal{A}_H$ are obtained from an expansion at the horizon. We further show that the solutions satisfy a Smarr formula.\\
We apply an ansatz based on the rod structure and parameterization in terms of canonical coordinates employed by \citet{Harmark:2004rm}. For the numerical construction of the black rings, however, it has turned out to be more suitable to reparameterize the ansatz in terms of isotropic coordinates.\\
In Sect.~\ref{sec:Action}, we will recall the Einstein-Maxwell action and the respective equations for the metric and the gauge potential. The ansatz for a stationary and axially symmetric solution as well as the boundary conditions will be given in Sect.~\ref{sec:Ansatz}. In Sect.~\ref{sec:Quantities}, the physical properies of the charged balanced black ring will be discussed. We present our numerical results in Sect.~\ref{sec:Results}, focussing on large thin black rings. A conclusion and an outlook will be given in Sect.~\ref{sec:Conclusions}.

\section{EINSTEIN-MAXWELL ACTION}
\label{sec:Action}
In order to obtain electrically charged black ring solutions of pure Einstein-Maxwell theory, we consider the five-dimensional Einstein-Maxwell action
\begin{equation}
S=\frac{1}{16\pi G_5}\int d^5 x\sqrt{-g}(R-F_{ab}F^{ab})
\label{eq:Action}
\end{equation}
with the curvature scalar $R$, the five-dimensional Newton constant $G_5$ and the field strength tensor $F_{ab}=\partial_a A_b-\partial_b A_a$, where $A_a$ denotes the gauge potential.\\

A variation with respect to the metric tensor leads to the Einstein equations
\begin{equation}
G_{ab}=2 T_{ab},
\label{eq:EM-eq}
\end{equation}
where the Einstein tensor $G_{ab}$ and the energy-stress tensor $T_{ab}$ are given by:
\begin{equation}
G_{ab}=R_{ab}-\frac{1}{2}R g_{ab}
\end{equation} 
and
\begin{equation}
T_{ab}=F_{ac}F_{b}^{c}-\frac{1}{4}g_{ab}F_{cd}F^{cd}.
\end{equation} 
The Maxwell equations are obtained by varying the action with respect to the gauge potential:
\begin{equation}
\nabla_{b}F^{ab}=0.
\label{eq:M-eq}
\end{equation}

\section{ANSATZ AND BOUNDARY CONDITIONS}
\label{sec:Ansatz}
The approach for the stationary axially symmetric black ring metric is based on the metric in canonical coordinates $\rho,z$ and the rod structure used by \citet{Harmark:2004rm}, with the  semi-infinite spacelike rod $\rho=0, -\infty\leq z\leq -a$ in $\psi$-direction, the finite timelike rod $\rho=0, -a\leq z\leq a$, the finite spacelike rod $\rho=0, a\leq z\leq b$ in $\psi$-direction and the semi-infinite spacelike rod $\rho=0, b\leq z\leq \infty$ in $\varphi$-direction.\\ 
Since the quantity $a$ determines the length of the finite time-like rod, it can be related to the size of the $S^2$. The quantity $b$, fixing the length of the finite $\psi$-rod, can roughly be seen as a measure of the radius of the $S^1$ \citep{Kleihaus:2009wh, Kleihaus:2009dm}, with large thin black rings corresponding to $2<b$ and small fat black rings being obtained for $1<b<2$.\\
\begin{figure}
\centering
\includegraphics[width=8cm]{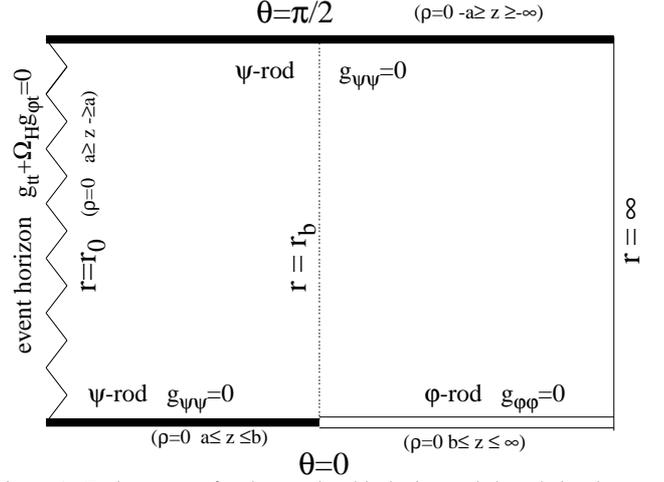}
\caption{Rod structure for the rotating black ring and the relation between isotropic and canonical coordinates}
\label{fig:Rods}
\end{figure}
For the numerical calculations, we have found it to be more suitable to reparameterize the metric in terms of isotropic coordinates $r,\theta$ according to
\begin{equation}
\rho=\frac{r^4-r_0^4}{2r^2}\mbox{sin}2\theta, \qquad z=\frac{r^4+r_0^4}{2r^2}\mbox{cos}2\theta,
\end{equation}
with $r_0^2=a$ being the horizon radius of the $S^2$. This yields the following ansatz for the metric:
\begin{eqnarray}
ds^2&=&-f_0(r,\theta) dt^2+\frac{1}{f_1(r,\theta)}(dr^2+r^2 d\theta^2)\\
& &+f_2(r,\theta)d\psi^2+f_3(r,\theta)(d\varphi-\frac{\omega(r,\theta)}{r}dt)^2.\nonumber
\label{eq:metricrot_iso}
\end{eqnarray}
Here the cross term in $d\varphi dt$ represents a rotation in $\varphi$-direction, hence a rotation along the $S^1$. All metric functions depend on $r$ and $\theta$ only.\\
The relation between isotropic and canonical coordinates is depicted in Fig.~\ref{fig:Rods}.
In isotropic coordinates, the horizon is mapped to $r=r_0,0\leq\theta\leq\frac{\pi}{2}$ and infinity to $r\rightarrow\infty,0\leq\theta\leq\frac{\pi}{2}$. The semi-infinite and the finite $\psi$-rod are located at $r_0\leq r\leq\infty,\theta=\frac{\pi}{2}$ and $r_0\leq r\leq r_b,\theta=0$ respectively. The semi-infinite $\varphi$-rod is mapped to $r_b\leq r\leq\infty,\theta=0$, with $r_b=\sqrt{b+\sqrt{b^2-a^2}}$. Thus, the boundaries are all orthogonal to each other except for the finite $\psi$-rod and the semi-infinite $\varphi$-rod intersecting at $\theta=0,r=r_b$.\\
The ansatz for the gauge potential is given by
\begin{equation}
A_{a}(r,\theta)dx^{a}=A_0(r,\theta)dt+A_{\varphi}(r,\theta)d\varphi.
\end{equation}
%the $\varphi$-component arising due to the rotation.\\

% The equations for the metric functions $f_i(r,\theta)$ and the gauge potential $A(r,\theta )$ are found by using a suitable combination of the Einstein-Maxwell equations for the ansatz Eq.~\ref{eq:metricrot_iso}, i.e.
% \begin{equation}
% G_t^t=2 T_t^t,\ G_r^r+G_{\theta}^{\theta}=2(T_r^r+T_{\theta}^{\theta}),\ G_{\psi}^{\psi}=2T_{\psi}^{\psi},\ G_{\varphi}^{\varphi}=2T_{\varphi}^{\varphi},\ G_{t}^{\varphi}=2T_t^{\varphi},
% \end{equation} 
% as well as the Maxwell equations
% \begin{equation}
% \nabla_{b} F^{tb}=0,\quad \nabla_{b} F^{\varphi b}=0.
% \end{equation}
% The remaining Einstein-Maxwell equations $G_{\theta}^r=2T_{\theta}^r$ and $G_r^r-G_{\theta}^{\theta}=2(T_r^r-T_{\theta}^{\theta})$ deliver two further constraints \citep{Kleihaus:2009wh, Kleihaus:2009dm}.
% The non-vanishing components of $G_{ab}$ and $T_{ab}$ are shown in App.~\ref{app:C}.\\

At infinity, the metric functions should approach the Minkowski metric, hence we impose the boundary conditions $f_0=f_1=f_2=f_3=1$, $\omega=0$ for $r\to\infty$. The gauge potential must satisfy $A_0=A_{\varphi}=0$ at infinity by choice of gauge.\\
On each rod, one of the functions $f_0$, $f_2$, $f_3$ becomes zero, while the other $f_i$ stay finite. The event horizon is represented by the finite timelike rod, characterized by the condition $f_0(r_0)=0$. Here the other metric functions fulfill $\partial_{r}f_1=\partial_{r}f_2=\partial_{r}f_3=0$ and $\omega=\omega_H$, with $\Omega_H=\omega_H/r_0$ representing the horizon angular velocity. The gauge field satisfies $A_0+\Omega_H A_{\varphi}=-\Phi_H$ and $\partial_r A_{\varphi}=0$ on the horizon.\\
On the $\psi$-rods we impose $f_2=0$ as well as $\partial_{\theta}f_0=\partial_{\theta}f_1=\partial_{\theta}f_3=\partial_{\theta}\omega=0$ for the metric functions and $\partial_{\theta}A_0=\partial_{\theta} A_{\varphi}=0$ for the gauge potential. The boundary conditions on the semi-infinite $\varphi$-rod are given by $f_3=0$,  $\partial_{\theta}f_0=\partial_{\theta}f_1=\partial_{\theta}f_2=\partial_{\theta}\omega=0$ and $\partial_{\theta}A_0=0$, $A_{\varphi}=0$.

% The boundary conditions are $f_0(r,\theta)\left.\right|_{r=r_0}=0$ and $\left.\partial_{r} f_i(r,\theta)\right|_{r=r_0}=0$ ($i\neq 0$) on the horizon, $f_2(r,\theta)\left.\right|_{\theta=\pi/2}=0$ and $\left.\partial_{\theta} f_i(r,\theta)\right|_{\theta=\pi/2}=0$ ($i\neq 2$) on the semi-infinite $\psi$-rod, $f_2(r,\theta)\left.\right|_{\theta=0,r\leq r_b}=0$ and $\left.\partial_{\theta} f_i(r,\theta)\right|_{\theta=0,r\leq r_b}=0$ ($i\neq 2$) on the finite $\psi$-rod and $f_3(r,\theta)\left.\right|_{\theta=0,r\geq r_b}=0$ and $\left.\partial_{\theta} f_i(r,\theta)\right|_{\theta=0,r\geq r_b}=0$ ($i\neq 3$) on the semi-infinite $\varphi$-rod.
% 
% The boundary conditions for the function $\omega(r,\theta)$ are given by
% $\omega(r,\theta)=\mbox{const.}=\omega_H$ on the horizon (with $\Omega_H=\omega_H/r_0$ being the horizon angular velocity), $\partial_{\theta}\omega(r,\theta)=0$ on the $\psi$-rods and $\varphi$-rod and $\omega(r,\theta)=0$ at infinity. At infinity, the metric functions should approach the Minkowski metric
% \begin{equation}
% ds^2=-dt^2+dr^2+r^2(d\theta^2+\mbox{cos}^2\theta d\psi^2+\mbox{sin}^2\theta d\varphi^2).
% \label{eq:Minkowski}
% \end{equation}
% 
% For the gauge potential, the boundary conditions $A_0+\Omega_H A_{\varphi}=-\Phi_H, \partial_r A_{\varphi}=0$ on the horizon, $\partial_{\theta}A_0=0, \partial_{\theta} A_{\varphi}=0$ on the $\psi$-rods,  $\partial_{\theta}A_0=0, A_{\varphi}=0$ on the $\varphi$-rod and $A_0=0, A_{\varphi}=0$ at infinity have to be satisfied.\\

For the numerical calculations the metric functions $f_1$, $f_2$ and $f_3$ are expressed as a product
\begin{equation}
f_i=f_i^0F_i,
\end{equation}
of background functions $f_i^0$ times some functions $F_i$ displaying the deviation from the respective background function. Here we choose the background functions as the respective metric functions of the neutral static black ring, obtained by a transformation of the respective metric functions in canonical coordinates \citep{Harmark:2004rm} to isotropic coordinates.\\
By this choice, the desired rod structure is automatically fulfilled and discontinuities at $r_b$ as well as divergences of the functions $f_2$ and $f_3$, coming from the imposed asymptotic behaviour, are absorbed. For the functions $f_0$, $\omega$, $A_0$ and $A_{\varphi}$ no background functions are introduced.
The boundary conditions for the $F_i$ are $F_i=1$ at infinity, $\partial_r F_i=0$ on the horizon and $\partial_{\theta}F_i=0$ along the rods, with the exception of $F_1F_3=1$ along the semi-infinite $\varphi$-rod, following from the requirement of regularity.

\section{THE PHYSICAL PROPERTIES}
\label{sec:Quantities}
The horizon metric is given by
\begin{equation}
ds_H^2=\frac{1}{f_1(r_0,\theta)}r_0^2 d\theta^2+f_2(r_0,\theta)d\psi^2+f_3(r_0,\theta)d\varphi^2,
\end{equation} 
so the horizon area $\mathcal{A}_H$ of the black ring is calculated according to
\begin{equation}
\mathcal{A}_H=(2\pi)^2 r_0\int_{0}^{\pi/2} d\theta \sqrt{\frac{f_2(r_0,\theta)f_3(r_0,\theta)}{f_1(r_0,\theta)}}.
\end{equation}
Here we have chosen $\Delta\psi=2\pi$, so the solution is asymptotically flat. By this choice, unbalanced solutions contain a conical singularity for the finite $\psi$-rod, in which case the ring is sitting on the rim of a disk-like deficit membrane preventing it from collapsing.\\
The Hawking temperature $T_H$ is obtained by setting $t=i\tau$ and requiring regularity on the Euclidean section \citep{Kleihaus:2009dm}:
\begin{equation}
T_H=\frac{1}{2\pi}\lim_{r\to r_0}\sqrt{\frac{f_0(r,\theta)f_1(r,\theta)}{(r-r_0)^2}},
\end{equation}
while the conical singularity $\delta$ on the finite $\psi$-rod is calculated according to \citep{Kleihaus:2009dm}
\begin{equation}
\delta=2\pi-\lim_{\theta\to 0}\sqrt{\frac{f_2(r,\theta)f_1(r,\theta)}{r^2\theta^2}}.
\end{equation}

The mass $M$ and the angular momentum $J$ of the solution can be read off at infinity via the asymptotic expansion of the metric \citep{Kunz:2005ei}
\begin{equation}
f_0\to -1+\frac{8G_5 M}{3\pi}\frac{1}{r^2},\qquad\omega\to \frac{4G_5 J}{\pi}\frac{1}{r^3},
\end{equation}
whereas the charge $Q$ and the magnetic moment $\mathcal{M}_{\varphi}$ are obtained from the asymptotic expansion of the gauge potential
\begin{equation}
A_0\to -\frac{G_5 Q}{\pi}\frac{1}{r^2},\qquad A_{\varphi}\to \frac{G_5\mathcal{M}_{\varphi}\mbox{sin}^2\theta}{\pi r^2}.
\end{equation}
The gyromagnetic ratio is given by
\begin{equation}
g=\frac{2M\mathcal{M}_{\varphi}}{QJ}.
\end{equation}

In order to analyze the physical properties of the solution, it is convenient to work with dimensionless quantities, obtained by dividing by an appropriate power of $M$ or $G_5 M$. This gives the scaled horizon area $a_H$ and the scaled temperature $t_H$ 
\begin{equation}
a_H=\frac{3}{16}\sqrt{\frac{3}{\pi}}\frac{\mathcal{A}_H}{(G_5 M)^{3/2}},\qquad t_H=T_H\sqrt{G_5 M},
\end{equation}
as well as the scaled squared angular momentum $j^2$ and the scaled charge $q$
\begin{equation}
j^2=\frac{27\pi}{32 G_5}\frac{J^2}{M^3},\qquad q=\frac{Q}{M}.
\end{equation}

\section{NUMERICAL RESULTS}
\label{sec:Results}
\begin{figure}
\centering
\includegraphics[width=8cm]{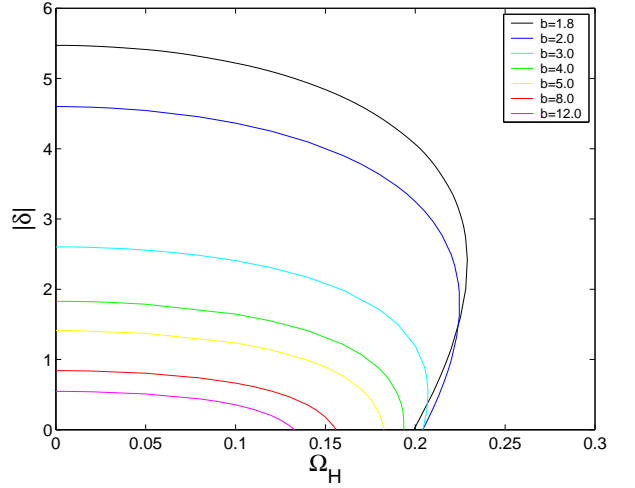}
\caption{Dependence of the conical singularity $\delta$ on the parameters $b$ and $\Omega_H$ for neutral black rings ($\alpha=0$)}
\label{fig:delta_uncharged}
\end{figure}
\begin{figure}
\centering
\includegraphics[width=8cm]{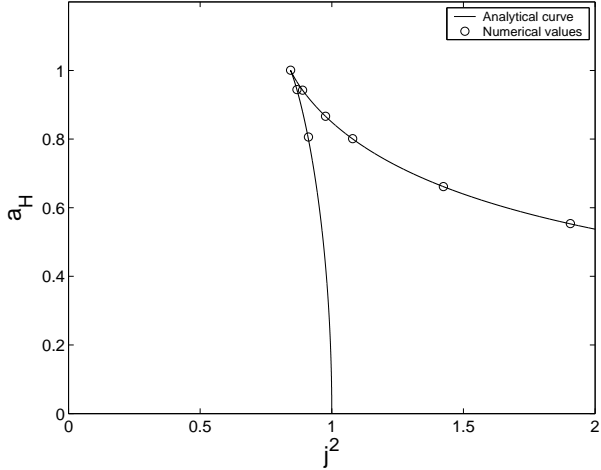}
\caption{Scaled horizon area $a_H$ versus scaled squared angular momentum $j^2$ for the neutral rotating black ring (see \citep{Emparan:2006mm, Emparan:2008eg} for the analytical curve)}
\label{fig:ja_uncharged}
\end{figure}
For the numerical calculations, we introduce the compactified radial variable $x=1-r_0/r$, which maps the semi-infinite region $[r_0,\infty]$ to the finite region $[0,1]$. The resulting system of seven coupled non-linear elliptic partial differential equations is solved numerically with the help of the finite difference solver FIDISOL, based on the Newton-Raphson method.\\
The parameters of the solution are given by the positions of the finite $\psi$-rod $a$ and $b$ ($r_0$ and $r_b$, respectively), the charge parameter $\alpha=\sqrt{4/3}\cdot\Phi_H$ and the rotation parameter $\Omega_H$. We fix $r_0=1$ for all solutions and vary the other parameters.\\
In the limit of $\Omega_H=0$, we observe excellent agreement with the analytical solution by \citet{Yazadjiev:2005hr}, with a maximum deviation of order $10^{-5}$ only. 
%while for $\alpha=0$ we reproduce the neutral rotating black ring solution by \citet{Emparan:2001wn}.\\
In the static case as well as for generic values of the rotation parameter, the solutions suffer from a conical deficit. However for particular parameter values, we have obtained regular solutions whose horizons are smooth, where tension and gravitational self-attraction are balanced by the repulsion due to the rotation and the electric charge of the ring.\\
We find that the conical singularity $\delta$ decreases with increasing $\alpha$ as well as with increasing $b$ and $\Omega_H$. For given values of $\alpha$ and $b$, we vary $\Omega_H$, until the balanced solution is found. An alternative way for obtaining balanced solutions would be to increase the charge parameter for a fixed value of the rotation parameter. The rotation parameter however exhibits greater influence on the conical singularity.\\ 
We have obtained balanced solutions for the neutral case first. As we increase $\Omega_H$ from zero for fixed values of $\alpha$ and $b$, from the corresponding static black ring a branch of rotating solutions emerges, generically exhibiting a conical deficit. This upper branch extends up to a maximum value for $\Omega_H$, depending both on $\alpha$ and $b$. A second lower branch bends back from the maximal value of $\Omega_H$ towards $\Omega_H=0$. Along both branches, the mass $M$, the angular momentum $J$ and the charge $Q$ rise monotonically. In Fig.~\ref{fig:delta_uncharged}, the dependence of the conical deficit on the rotation parameter $\Omega_H$ is shown for $\alpha=0$ and $1.8\leq b\leq 12.0$. While for small values of the parameter $b$ the balanced solution is located on the lower branch, for larger values of $b$ the balanced solution is already obtained before the maximum value of $\Omega_H$ is reached.\\
In Fig.~\ref{fig:ja_uncharged}, we exhibit the scaled horizon area $a_H$ versus the scaled squared angular momentum $j^2$, showing that our numerically obtained balanced solutions agree excellently with the analytical solution by \citet{Emparan:2001wn}, with deviations of order $10^{-4}$ only (see e.g.~\citep{Emparan:2006mm, Emparan:2008eg} for the analytical phase diagram).\\

\begin{figure*}
\centering
\includegraphics[width=18cm]{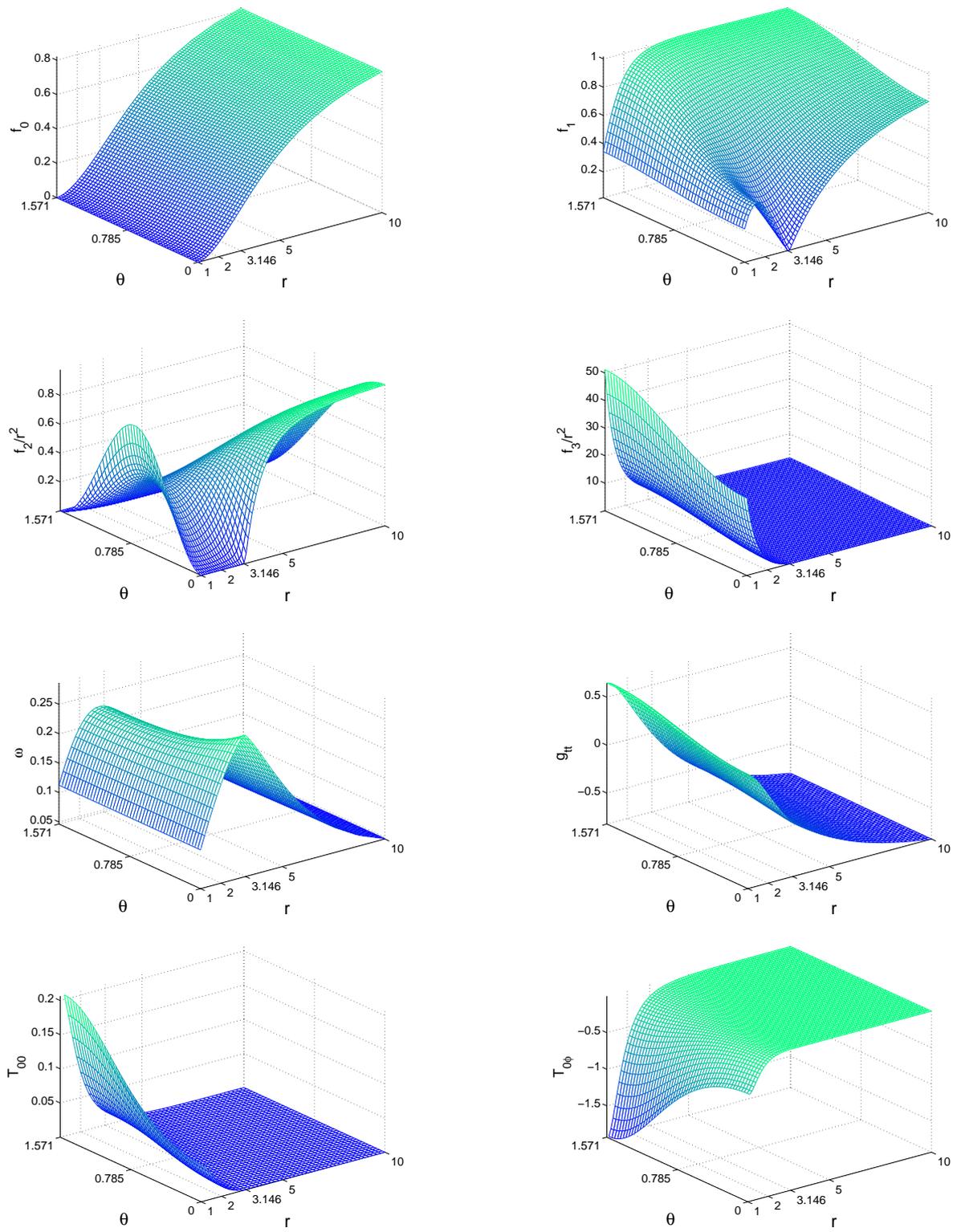}
\caption{Metric functions, $T_{00}$-component and $T_{0\varphi}$-component of the energy-stress tensor for the balanced solution belonging to $b=5,\Phi_H=\sqrt{3/4}\cdot0.6,Q=22.02,J=49.35$}
\label{fig:Solution}
\end{figure*}

The metric functions, the $T_{00}$-component as well as the $T_{0\varphi}$-component of the energy-stress tensor $T_{ab}$ are shown in Fig.~\ref{fig:Solution} for the charged balanced solution characterized by $b=5,\Phi_H=\sqrt{3/4}\cdot0.6,Q=22.02,J=49.35$.\\
\begin{figure}
\centering
\includegraphics[width=8cm]{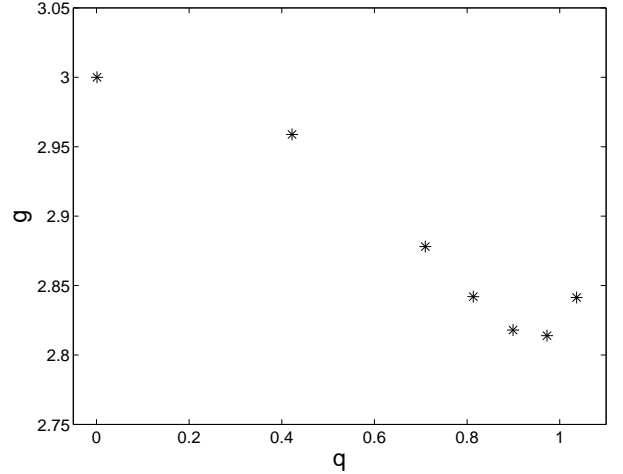}
\caption{Dependence of the gyromagnetic ratio $g$ on the scaled charge $q$ for charged balanced black ring solutions belonging to $b=5$}
\label{fig:gFaktor}
\end{figure}
For the charged case, we have so far obtained balanced solutions for a set of large thin black rings, namely for $b=3$, $b=5$, $b=8$ and $b=12$. Since the numerical calculations are more difficult on the lower branch and in the transition region between the two branches, it gets increasingly demanding to obtain solutions for small values of $b$ and hence for rings with decreasing radius of the $S^1$. A full exploration of the parameter space is beyond the scope of this letter. \\

Calculating the physical quantities, we find that our solutions satisfy the Smarr relation \citep{Kunz:2005ei, Gauntlett:1998fz} (with deviations after five digits only): 
\begin{equation}
M=\frac{3}{16\pi G_5}\kappa \mathcal{A}_H+\frac{3}{2}\Omega_H J+\Phi_H Q,
\label{eq:Smarr_rot}
\end{equation}
where $\kappa=2\pi T_H$ is the surface gravity.\\

The Hawking temperature $T_H$ is found to be constant on the horizon (with deviations after 6 digits only), as required by the zeroth law of black hole mechanics. Both the scaled temperature $t_H$ and the scaled horizon area $a_H$ decrease with increasing scaled charge $q$. The magnetic moment $\mathcal{M}_\varphi$ rises with increasing scaled charge. In Fig.~\ref{fig:gFaktor} we exhibit the gyromagnetic ratio $g$ versus the scaled charge $q$ for balanced solutions belonging to $b=5$. We have found $g=D-2=3$ for weakly charged rings, in accordance with \citet{Ortaggio:2006ng}. For increasing values of the scaled charge however, a negative deviation of $g$ from this value can be observed, similar to the case of rotating electrically charged black holes with a spherical horizon topology and a single angular momentum \citep{Kunz:2005nm}.\\

\begin{figure}
\centering
\includegraphics[width=8cm]{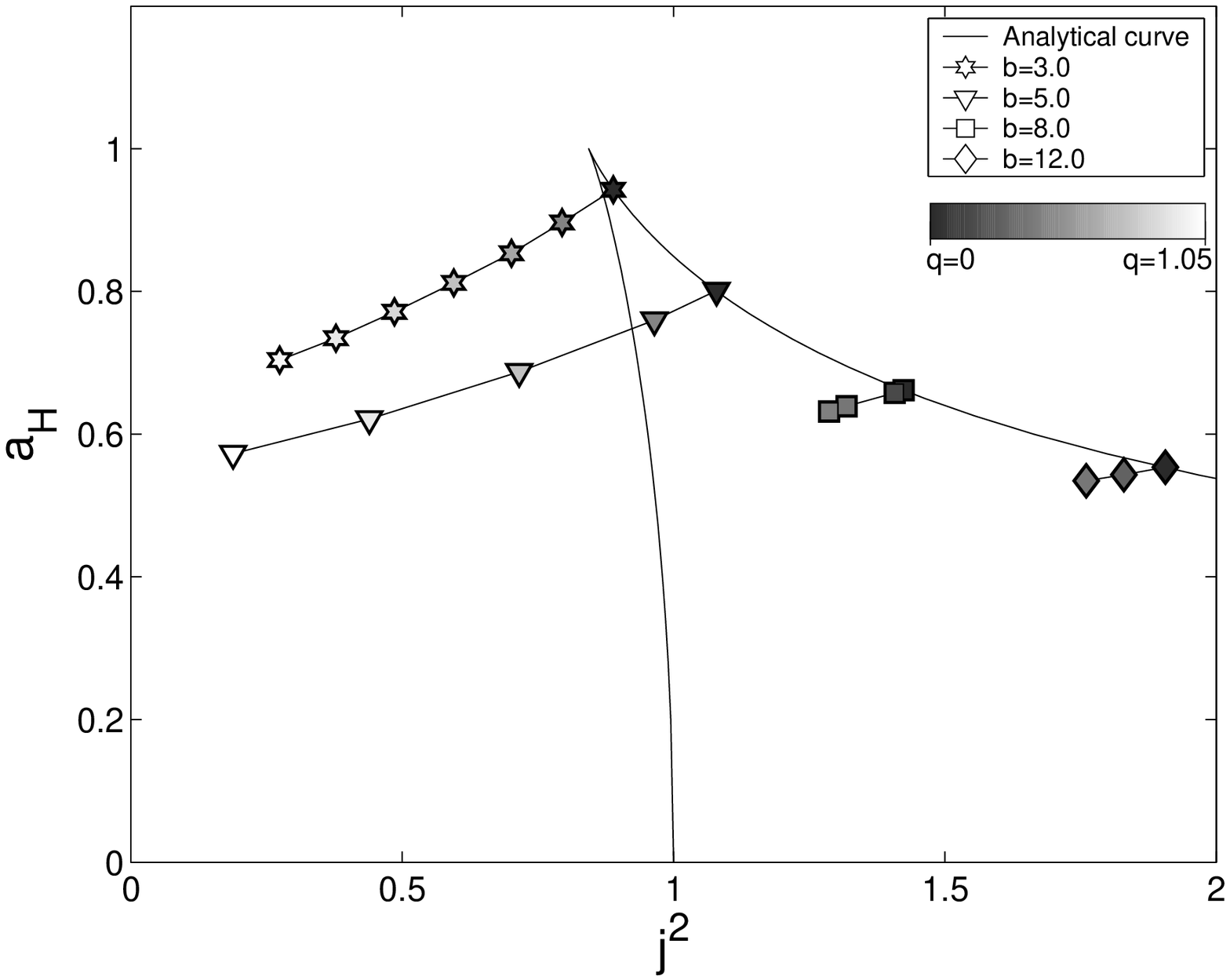}
\caption{Scaled horizon area $a_H$ versus scaled squared angular momentum $j^2$ for the neutral rotating black ring (see \citep{Emparan:2006mm, Emparan:2008eg} for the neutral analytical curve) and for the electrically charged balanced black ring}
\label{fig:ja_+charged}
\end{figure}
In Fig.~\ref{fig:ja_+charged}, we exhibit the scaled horizon area $a_H$ versus the scaled squared angular momentum $j^2$ and the scaled charge $q$ for the charged balanced numerical solutions together with the neutral analytical solution by \citet{Emparan:2001wn}. We observe that an increasing value of the scaled charge $q$ makes the conical deficit vanish for lower values of the rotation parameter $\Omega_H$  and hence lower values of the scaled squared angular momentum $j^2$. The non-scaled angular momentum, however, increases with increasing charge. The complete phase diagram is under construction and will be given elsewhere.\\
\begin{figure}
\centering
\includegraphics[width=8cm]{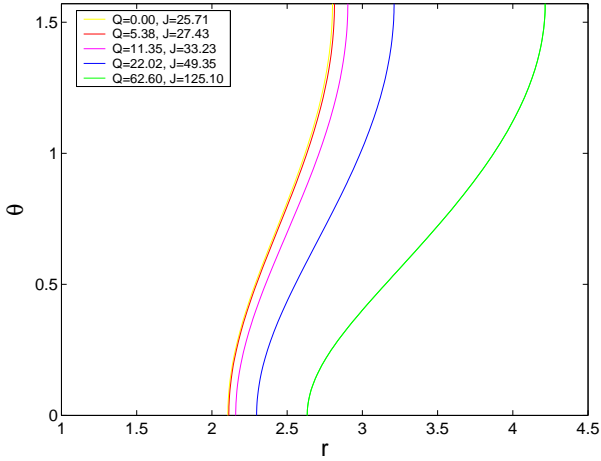}
\caption{Location of the ergosurface $g_{tt}=0$ (for $b=5$)}
\label{fig:Ergo}
\end{figure}
The location of the ergosurface $g_{tt}=0$ is depicted in Fig.~\ref{fig:Ergo}, showing that the ergosurface moves farther away from the horizon with increasing charge and angular momentum. 

\section{CONCLUSIONS}
\label{sec:Conclusions}
% In five-dimensional spacetimes, besides the spherical black hole solutions, black rings represent an additional, topologically different solution to the Einstein equations. While there have been found many explicit black ring solutions in various gravity theories and electrically charged rotating black rings have been derived in theories including additional fields, an analytical solution of pure Einstein-Maxwell theory is not known so far. In this work however, such a solution has been derived numerically.\\
% 
% After an introduction to the most important analytically known black ring solutions and their physical properties, the ansatz for finding the metric as well as the gauge potential of a charged rotating black ring have been presented. Here, the rotation has been restricted to occur along the $S^1$ of the ring. The extraction of the physically relevant quantities and the numerical methods have been described. Finally the numerical results have been presented, and the properties of this new solution have been discussed.\\

We have considered rotating black ring solutions of five-dimensional Einstein-Maxwell theory. The solutions are asymptotically flat and possess a single angular momentum associated with a rotation along the $S^1$ of the ring. Whereas our approach has been analytical with respect to the expansions at the horizon and at infinity, the actual construction of the solution has been done numerically. Here we have parameterized the metric in terms of isotropic coordinates which are well-suited for the numerical work.\\
In the static case as well as for generic values of the rotation parameter, the solutions suffer from a conical deficit. For particular parameter values, however, we have succeeded in obtaining regular solutions whose horizons are smooth and whose tension and gravitational self-attraction are balanced by a repulsion provided by the rotation and the charge of the ring.\\
Concerning the numerical accuracy, excellent agreement with the static analytical solution by \citet{Yazadjiev:2005hr} and the neutral analytical solution by \citet{Emparan:2001wn} has been found, with an accuracy of at least $10^{-5}$ and $10^{-4}$, respectively. 
% In the neutral limit, the rotating vacuum black ring solution by \citet{Emparan:2001wn} has been reproduced. 
All solutions satisfy a Smarr relation, indicating the high accuracy of the numerical results.\\ 
The solutions depend on four parameters. Having fixed the horizon radius ($r_0=1$), the solutions thus depend on the three remaining parameters. Those are the horizon angular velocity $\Omega_H$, the charge parameter $\alpha$ and the parameter $b$ representing a rough measure for the radius of the $S^1$.\\
The physical properties of the solution have been extracted and their dependence on the charge has been examined. As required by the zeroth law of black hole mechanics, the Hawking temperature $T_H$ is found to be constant on the horizon.\\
The conical singularity $\delta$ decreases both with growing values of the charge parameter $\alpha$  and the rotation parameter $\Omega_H$, and with increasing parameter $b$.\\ 
The scaled temperature $t_H$ and scaled horizon area $a_H$ decrease with increasing scaled charge $q$. When examining the functional dependence of the scaled angular momentum $j$ on the scaled charge $q$ for the regular solutions, the scaled angular momentum is found to decrease with increasing scaled charge. Thus solutions with higher scaled charge need less scaled angular momentum to be balanced.\\ 
While the gyromagnetic ratio for weakly charged rings has been found to be $g=3$, in accordance with \citet{Ortaggio:2006ng}, a negative deviation from this value has been observed for larger magnitudes of the electric charge similar to the case of black holes with a spherical horizon topology \citep{Kunz:2005nm}.\\
The ergosurface of the balanced solutions moves farther away from the horizon with increasing scaled charge.\\
Since it is numerically increasingly difficult to find balanced solutions for smaller values of the parameter $b$, i.e.~a decreasing size of the $S^1$, the parameter space has not been fully explored yet. The complete phase diagram is under construction and will be presented elsewhere.\\
A next step will be to include also rotation along the $S^2$ of the ring, leading to two additional metric functions and one further non-vanishing component of the gauge potential in $\psi$-direction. An interesting further step should be the inclusion of a Chern-Simons $A F^2$ term into the action, modifying the equations for the gauge potential while leaving the Einstein equations unaltered. Such an approach yields supersymmetric black rings for a Chern-Simons coupling value of $\lambda_{CS}=1$, for which an analytical solution has been obtained by \citet{Elvang:2004rt}. Here the phase diagram shows uniqueness of the solutions. It will be interesting to see how the phase diagram depends on the value of the Chern-Simons coupling constant $\lambda_{CS}$, where so far only rotating black holes with spherical horizon topology have been constructed \citep{Kunz:2005ei}.\\
We hope that the numerical solutions presented in this work will be helpful for the analytical construction of the exact five-dimensional electrically charged rotating Einstein-Maxwell black ring solution.\\
{\bf{Acknowledgement:}} We gratefully acknowledge discussions with E.~Radu. B.~K.~acknowledges support by the DFG.


%% References


%\bibliography{BalRingsBib}





% \end{thebibliography}

\end{document}